\documentclass[aps,pra,amsmath,amssymb,amsfonts,superscriptaddress,lengthcheck,twocolumn,longbibliography]{revtex4-1}
\usepackage{dcolumn}    
\usepackage{graphicx}
\usepackage{amsmath, bbm}
\usepackage{latexsym}
\usepackage{amsfonts}   
\usepackage{amssymb}
\usepackage{array}      
\usepackage{epsfig}
\usepackage{txfonts}
\usepackage{xcolor,braket}
\usepackage[colorlinks=true,linkcolor=blue,urlcolor=blue,citecolor=blue,pdfusetitle]{hyperref}
\usepackage{hyperref}
\usepackage{ulem}
\usepackage{soul}
\usepackage{dsfont}
\usepackage{physics}
\usepackage{bm}
\usepackage{tensor}

\usepackage{relsize}

\usepackage{mathrsfs}

\usepackage{cancel}
\usepackage{enumitem}

\begin{document}
 \title{Fisher information rates in sequentially measured quantum systems} 
       
\author{Eoin O'Connor}
\affiliation{School of Physics, University College Dublin, Belfield, Dublin 4, Ireland}
\affiliation{Centre for Quantum Engineering, Science, and Technology, University College Dublin, Belfield, Dublin 4, Ireland}
\author{Steve Campbell}
\affiliation{School of Physics, University College Dublin, Belfield, Dublin 4, Ireland}
\affiliation{Centre for Quantum Engineering, Science, and Technology, University College Dublin, Belfield, Dublin 4, Ireland}
\affiliation{Dahlem Center for Complex Quantum Systems, Freie Universit\"at Berlin, Arnimallee 14, 14195 Berlin, Germany}
\author{Gabriel T. Landi}
\affiliation{Department of Physics and Astronomy, University of Rochester, Rochester, New York 14627, USA}
\begin{abstract}
We consider the impact that temporal correlations in the measurement statistics can have on the achievable precision in a sequential metrological protocol. In this setting, and for a single quantum probe, we establish that it is the transitions between the measurement basis states that plays the most significant role in determining the precision, with the resulting conditional Fisher information being interpretable as a rate of information acquisition. Projective measurements are shown to elegantly demonstrate this in two disparate estimation settings. Firstly, in determining the temperature of an environment and, secondly, to ascertain a parameter of the system Hamiltonian. In both settings we show that the sequential estimation approach can provide a useful method to enhance the achievable precision.
\end{abstract}
\date{\today}

\maketitle

\section{Introduction}
Accurate measurements underpin our ability to understand all physical systems and processes. This places a high priority on the development of useful metrological protocols, i.e. estimation schemes to infer the maximal amount of information regarding some unknown parameters of a given system of interest. More formally, an estimation scheme is a process that converts measurement data into an estimate of an unknown parameter, $\theta$. The Cramér-Rao bound~\cite{cramer1999mathematical,rao1973} places a lower bound on the variance of any unbiased estimation scheme
\begin{equation}
    \sigma_\theta \geq \frac{1}{F_\theta},
    \label{eq::CR_into}
\end{equation} 
where $F_\theta$ is the associated Fisher information~\cite{fisher1922}. In standard quantum and classical estimation schemes the Fisher information scales linearly with the number of measurement results, $N$, leading to a $N^{-1}$ scaling in the variance. By making use of quantum correlations, such as entanglement, between different sub-systems it is possible for quantum estimation schemes to achieve $N^{-2}$ scaling, the so-called Heisenberg limit~\cite{Giovannetti_2004,Giovannetti_2006,Giovannetti_2011,Braun_2018}. It has recently been demonstrated that it may even be possible to surpass the Heisenberg limit by exploiting critical phenomena~\cite{FrerotPRL, Rams_2018,Garbe_2022,NimmrichterNew, KarolNew1, KarolNew2, GeorgeMPEst}.

In spite of the advantages, when dealing with quantum systems additional subtleties must be taken into account; such as the freedom in choosing the measurement basis or considering generalized measurement operators. In probe based metrology, where a sensor is placed in contact with a sample whose properties we wish to learn, measurement back-action can also play a significant role~\cite{braginsky1995quantum,busch1990quantum,wiseman1995using,gudder2001sequential,Naik_2006,Hatridge_2013}. In fact, even if the sample is sufficiently large that the interaction between it and probe has negligible impact on its state, such back-action can still play a significant role in the effectiveness of the protocol depending on how the measurements are performed on the probe.  If the {\it same} probe system is measured repeatedly, the backaction can cause each measurement result to depend on the outcomes of, in principle all, previous measurements. While the impact of this can be neglected if, as is often assumed, the probe is reset after each measurement, this has the deleterious effect of making the process slow and, more importantly, potentially throwing away the opportunity to extract a better estimation by using the additional information gained due to the correlations established between the outcomes. This is motivation behind sequential quantum metrological protocols~\cite{Nagali_2012,Kiilerich_2015,Burgarth_2015,Muller_2016,De_Pasquale_2017,bompais2023asymptotic}. Additionally, if the probe is reset between each measurement it also precludes us from exploiting any quantum correlations that may have built up in our system and results in the Fisher information of such processes necessarily scaling linearly in the number of repetitions. There has been recent numerical evidence that it is possible to achieve super-linear scaling as the number of repetitions of a sequential measurement process increases~\cite{Clark_2019, Montenegro_2022}. Although this scaling likely only holds for a finite number of repetitions~\cite{yang2023extractable}.

In the sequential setting the metrological process can be depicted as a time-series of outcomes, i.e. a stochastic process. Due to the measurement backaction, the outcomes will generally be correlated in time. These correlations can affect the rate at which we acquire information, however, as has recently been shown in Refs.~\cite{Radaelli_2023, Smiga2023}, these correlations are not always beneficial and can either speed up or slow down the acquisition of information. It is therefore highly relevant to critically assess how correlations in time impact the rate of learning and to develop schemes that leverage the measurement backaction to increase metrological precision.

In this work we take precisely this approach in probe-based metrology, where we consider a single system acting as the probe, which is interacting with a sample (hereafter referred to as the environment). The protocol involves performing sequential measurements on the probe at discrete time intervals that generate correlations between the measurements, which can be leveraged to increase the Fisher information. For a process with finite Markov order, the Fisher information will necessarily scale linearly in the long-time limit, in contrast to the large Markov order found, e.g. in continuous~\cite{Klaus1, Klaus2, Klaus3, Marco1, Marco2, Marco3, Marco4} and weak~\cite{Nori1, Nori3} measurement schemes, which can make parameter estimation difficult~\cite{radaelli2024parameter}. We demonstrate that stroboscopically performing projective measurements on the probe, whose underlying dynamics is otherwise Markovian, leads to a Markov order-1 measurement scheme. This results in a significantly simpler optimal estimation protocol and clear evidence of the role that temporal correlations in the measurement outcomes can have on the resulting Fisher information, which we can interpret as the rate of information acquisition. The small Markov order allows for effective feedback control~\cite{wiseman2009quantum,Zhang_2017} to be implemented simply by adjusting the time between measurements. We analyse this approach in two paradigmatic metrological settings, thermometry of a large environment and estimation of the Rabi frequency of a qubit.
 
\section{Correlated Fisher information}
We consider a general setup of a system, with initial state $\rho_S$ interacting with an environment at a state $\rho_E$, via a unitary $U$. From the perspective of the system, this can be  described by the completely positive trace-preserving (CPTP) map
\begin{equation}
    \mathcal{E}_\theta(\rho_S) = {\rm tr}_E\left\{ U (\rho_S \otimes \rho_E) U^\dagger\right\} = \sum_i K_i  \rho_S K_i^\dag,
    \label{eq::sysEvo}
\end{equation}
where $\{K_i\}$ are Kraus operators satisfying $\sum_i K_i^\dag K_i = \mathbf{I}$. In the most general setting, the state of the environment $\rho_E$ and/or the unitary $U$ depend on an unknown parameter $\theta$ which we wish to estimate. The channel $\mathcal{E}_\theta$ therefore transfers information about $\theta$ to the probe system's state. Such a setting encompasses several broad classes of dynamics, including the case of purely unitary dynamics of the system, which occurs when $U$ is a tensor product of unitaries. It also captures the case of generic quantum channels defined only by the Kraus operators, i.e. cases where we only have access to $\theta$-dependent $K_i$'s.

The basic task of probe-based metrology is to extract an estimate of $\theta$ from measurements of the system alone. That is, we perform a generalized measurement on the system described by a set of operators $L_\omega$ with outcomes $\omega$. The probability of obtaining outcome $\omega$ is 
\begin{equation}
    p(\omega|\rho_S) = \Tr(\rho_S  M_\omega),\qquad M_\omega = L_\omega^\dag L_\omega,
\end{equation}
and the state of the system, given that the outcome was $\omega$, is updated as 
\begin{equation}\label{eq::measurement_L}
    \rho_S \to \mathcal{L}_\omega(\rho_S)= \frac{L_\omega \rho_S L_\omega^\dag}{p(\omega|\rho_S)},
\end{equation}
where the channel $\mathcal{L}_\omega$ is now non-linear as it refers to a specific outcome $\omega$.

Typical probe based metrological protocols involve the resetting the state of the system and environment after each measurement, making each outcome $\omega$ iid (independent and identically distributed). The variance $\sigma_\theta$ of any estimator is bounded by the Cram\'{e}r-Rao bound
\begin{equation}
    \sigma_\theta \geq \frac{1}{N F_\theta^{\rm iid}},
    \label{eq::CRBound}
\end{equation}
where  $F_\theta^{\rm iid}$ is the Fisher information associated with a single outcome 
\begin{equation}
     F_\theta^{\rm iid} = \sum_{\omega\in\Omega} \frac{1}{p(\omega|\mathcal{E}_\theta(\rho_S))}\left(\frac{\partial}{\partial \theta}p(\omega|\mathcal{E}_\theta(\rho_S))\right)^2.
     \label{fisherinfo1}
\end{equation}
Resetting leads to the inevitable linear scaling for an iid measurement protocol. Relaxing the requirement to reset the system, and therefore moving beyond the iid setting, leads to correlations being established between the measurement outcomes and we examine the impact that these correlations have on the resulting Fisher information in the remainder of this Section.

\subsection{Correlated outcomes}
We consider a scenario where we sequentially measure the same probe~\cite{Gu_2011,cuatanua2012,Kiilerich_2015,Burgarth_2015,De_Pasquale_2017}, i.e. after each measurement is performed the same probe is made interact with the environment again. For ease of calculation, we assume that the environment is sufficiently large and therefore its state resets at each time as is the case for dynamics accurately captured under the Markov approximation, or when the environment is modelled by a suitable collision model~\cite{Ciccarello_2022,CampbellEPLReview}. In practice, this is tantamount to the assumption that the same channel $\mathcal{E}_\theta$ is applied each time. Crucially however, while the environment's state remains the same, the sequential nature of the process means that the system's state at the start of each interaction cycle will depend on the result of the previous measurements. Therefore, the results of the measurements are correlated with each other.  

We consider a process where we first measure the system according to Eq.~\eqref{eq::measurement_L}, then apply the channel~\eqref{eq::sysEvo} and then repeat this sequence. This leads to a string of outcomes $\omega_{1:N} := (\omega_1,\ldots,\omega_N)$. The state of the system conditioned on these $N$ outcomes will then be given by 
\begin{equation}\label{eq::state_system_N_outcomes}
    \rho_S(\omega_{1:N}) =  \mathcal{E}_\theta \circ \mathcal{L}_{\omega_N} \ldots \circ \mathcal{E}_\theta  \circ \mathcal{L}_{\omega_1} (\rho_S^0),
\end{equation}
where $\circ$ denotes map composition. At each step, the probability of obtaining the next outcome $\omega_{n+1}$ given all previous outcomes $\omega_{1:n}$ is 
\begin{equation}\label{eq::conditional_probability_general}
    p(\omega_{n+1}|\omega_{1:n}) = \Tr(\rho_S(\omega_{1:n}) M_{\omega_{n+1}}),
\end{equation}
and the probability of observing a particular sequence $\omega_{1:N}$ is 
\begin{equation}\label{general_stochastic_process}
    P(\omega_{1:N}) = p(\omega_N|\omega_{1:N-1}) \ldots p(\omega_2|\omega_1)p(\omega_1).
\end{equation}
This therefore describes a correlated (and generally non-Markov) stochastic process. 

The measure-evolve-repeat sequence provides sufficient versatility that we can naturally introduce a feedback mechanism where the applied channels $\mathcal{E}_\theta$ are assumed to be conditioned on the previous measurement outcome. This modifies Eq.~\eqref{eq::state_system_N_outcomes} according to 
\begin{equation}\label{eq::state_system_N_outcomes_feedback}
    \rho_S(\omega_{1:N}) = \mathcal{E}_\theta^{\omega_N} \circ \mathcal{L}_{\omega_N} \ldots \circ \mathcal{E}_\theta^{\omega_1}  \circ \mathcal{L}_{\omega_1} (\rho_S^0).
\end{equation}
This feedback could be introduced, e.g., by assuming that the unitary $U$ applied in Eq.~\eqref{eq::sysEvo} at each step depends on the previous measurement outcome. This can lead to more information about the parameter of interest allowing to increase the estimation precision over iid protocols, as we detail below and demonstrate with explicit examples.

To formalise our ideas, consider the map $\Phi_\omega(\rho)$ that represents one iteration of the sequential measurement process with a specific measurement outcome,
\begin{equation}\label{Phi}
    \Phi_\omega(\rho_S) = \mathcal{E}_\theta^\omega\circ \mathcal{L}_\omega \rho_S.
\end{equation}
We assume that the unconditional channel $\Phi(\rho_S)\!\equiv\!\sum_\omega p(\omega|\rho_S) \Phi_\omega(\rho_S) $ has a unique steady-state, $\Phi(\pi)\!=\!\pi$. 
We remark that $P(\omega_{1:N})$ is still conditioned on the initial state of the system.
However, as will become clear below, the choice of initial state plays only a small role in the sequential measurement scheme, and therefore to simplify the dynamics we assume the probe's initial state is given by the steady state; i.e., $\rho_S^0 = \pi$, thus making the probability distribution $P(\omega_{1:N})$ ``stationary''~\cite{Radaelli_2023}.

Since the resulting stochastic process~\eqref{general_stochastic_process} is now correlated, the Cram\'er-Rao bound~\eqref{eq::CRBound} becomes 
\begin{equation}
    \sigma_\theta \geq \frac{1}{F_\theta(\omega_{1:N})},
\end{equation}
where 
\begin{equation}\label{eq::F_general}
    F_\theta(\omega_{1:N}) = \sum_{\omega_1,\ldots,\omega_N} \frac{1}{P(\omega_{1:N})}\left(\frac{\partial}{\partial \theta}P(\omega_{1:N})\right)^2.
\end{equation}
The computation of $F_\theta(\omega_{1:N})$ is, in general, quite difficult as it involves a high-dimensional summation. Recently, it was shown by some of us~\cite{Radaelli_2023} that 
the calculation of Eq.~\eqref{eq::F_general} simplifies for processes having a finite Markov order $\mathcal{M}$, an assumption which is true in many cases of interest. It is also approximately true in cases with infinite Markov order, as one can often define some sufficiently high effective Markov order $\mathcal{M}_{\rm eff}$~\cite{yang2023extractable}. For systems with a finite Markov order Eq.~\eqref{eq::F_general} reduces to 
\begin{equation}\label{eq::Fisher_decomposition}
    F_\theta(\omega_{1:N}) = F_\theta(\omega_{1:\mathcal{M}}) + (N-\mathcal{M}) F_{\theta}(\omega_{\mathcal{M}+1}|\omega_{1:\mathcal{M}}).
\end{equation}
The first term is the Fisher information of a block of $\mathcal{M}$ outcomes, while the second is the conditional Fisher information, defined as
\begin{equation}\label{Fisher_conditional}
    F_{\theta}(\omega_{\mathcal{M}+1}|\omega_{1:\mathcal{M}})=
    \sum_{\omega_1,\ldots,\omega_{\mathcal{M}}}P(\omega_{1:\mathcal{M}}) \sum_{\omega_{\mathcal{M}+1}} 
    \frac{\left(\partial_\theta p(\omega_{\mathcal{M}+1}|\omega_{1:\mathcal{M}})\right)^2}{p(\omega_{\mathcal{M}+1}|\omega_{1:\mathcal{M}})}.
\end{equation}
The quantity $p(\omega_{\mathcal{M}+1}|\omega_{1:\mathcal{M}})$ is the probability of future outcomes given all the relevant past, i.e. up to the Markov order. Equation~\eqref{Fisher_conditional} is the Fisher information of this distribution averaged over all possible pasts. For $N \gg \mathcal{M}$, Eq.~\eqref{eq::Fisher_decomposition} shows that the dominant contribution to the Fisher information is given by $F_\theta(\omega_{1:N}) \simeq N F_{\theta}(\omega_{\mathcal{M}+1}|\omega_{1:\mathcal{M}})$. This therefore allows us to interpret the conditional Fisher information as a Fisher information rate
\begin{equation}\label{Fisher_information_rate_general}
    F_{\theta}(\omega_{\mathcal{M}+1}|\omega_{1:\mathcal{M}}) = \lim\limits_{N\to\infty} \frac{F_\theta(\omega_{1:N})}{N}.
\end{equation}
That is, it represents the effective Fisher information acquired per outcome. Notice that this is generally different from $F_\theta^{\rm iid}$, as defined in Eq.~\eqref{fisherinfo1}. In fact, as shown in Ref.~\cite{Radaelli_2023}, there is no general relation between the two quantities, and $F_{\theta}(\omega_{\mathcal{M}+1}|\omega_{1:\mathcal{M}})$ can be both smaller or larger than $F_\theta^{\rm iid}$ depending on the problem in question. 

\subsection{Projective measurements} 
\label{ssec:projective}
A particularly elegant and useful instance of this corresponds to when the measurement operators $L_\omega$ in Eq.~\eqref{eq::measurement_L} are projective measurements onto some basis $\{\ket{k}\}$. 
Since the projection erases all information about previous states, this corresponds to a Markov order $\mathcal{M}=1$; that is, each measurement result only depends on the previous outcome, i.e. when the underlying dynamics adheres to the Markov approximation. The probability of obtaining each outcome reduces to $p(k|\rho_S)\!=\! \langle k |\rho_S |k \rangle$ and the channel $\Phi_\omega(\rho_S)$ in Eq.~\eqref{Phi} simplifies to 
\begin{equation}
    \Phi_k(\rho_S) = \mathcal{E}_\theta^k \big(|k\rangle\langle k|\big).
\end{equation}
Hence, the conditional probability~\eqref{eq::conditional_probability_general} reduces to 
\begin{equation}\label{transition_probabilities}
    P(k|k') = \langle k | \Big[\mathcal{E}_\theta^{k'} \big(|k'\rangle\langle k'|\big) \Big] |k\rangle.
\end{equation}
This formula elegantly encompasses the relationship between the quantum channel $\mathcal{E}_\theta$ and the actual measurement record that is observed, which in this case has the form of a Markov chain. The steady-state of $\Phi_k$ is $\pi\!=\! \sum_k q_k |k\rangle\langle k|$ where $q_k$ is the solution of the Markov equation 
\begin{equation}\label{Markov_chain_steady_state_equation}
    q_k = \sum_{k'} q_{k'} P(k|k').
\end{equation}

For Markov order-1 processes, Eq.~\eqref{eq::Fisher_decomposition} simplifies to~\cite{Radaelli_2023}
\begin{equation}
    F_\theta(k_1,\ldots,k_N) = F_1 + (N-1)F_{2|1},
    \label{eq:seqFI2}
    \end{equation}
where we introduce the slightly simpler notation $F_1 \!=\!\sum_k (\partial_\theta q_k)^2/q_k$ for the Fisher information of the steady-state distribution $q_k$, and
\begin{equation}
    F_{2|1} = \sum_{k'} q_{k'} \sum_{k} \frac{1}{P(k|k')} \left(\frac{\partial}{\partial \theta} P(k|k')\right)^2,
    \label{eq::F21}
\end{equation}
for the conditional Fisher information. This quantity is precisely the Fisher information rate in Eq.~\eqref{Fisher_information_rate_general}. Interestingly, this result shows that for projective measurements, the rate at which we acquire information is directly related to the Fisher information of the transition probabilities $P(k|k')$. For a Markov order 1 process, we learn about $\theta$ by observing the transitions. It is therefore clear that the correlations between measurement outcomes in a sequential protocol will have an impact on the attainable precision. In Sec.~\ref{sec:comparisons} we demonstrate that these correlations can both enhance and hinder an estimation scheme, and subsequently in Sec.~\ref{ApplicationsSec} provide explicit examples of how they can be leveraged to boost the effectiveness of a given protocol.

\section{Comparison to other strategies}
\label{sec:comparisons}
The key insight arising from Eq.~\eqref{eq:seqFI2} is that for a metrological scheme employing sequential measurements on a single probe system, what matters for the acquisition of information are the {\it transitions} from $k'\to k$. This is clear from the fact that $F_{2|1}$ depends on $\partial_\theta P(k|k')$, i.e. on how sensitive the transitions are to changes in $\theta$. Conversely, $F_1$ depends on $\partial_\theta q_k$. One would naturally be tempted to compare $F_{2|1}$ with $F_1$, or to any other meaningful quantity. It turns out, however, that these comparisons are quite subtle and can, in fact, lead to incomplete or incorrect conclusions due to neglecting specific aspects of a given implementation. We now attempt to clarify this issue. 

\subsection{Comparing with $F_\theta^{\rm iid}$ in Eq.~\eqref{fisherinfo1}}
\label{ssec:comparing_iid}
A first, somewhat naive, choice would be to compare $F_{2|1}$ with the case where the outcomes are iid. 
There are two possible ways one might obtain iid outcomes. The first is to have $N$ copies of the probe and send each one individually through the channel $\mathcal{E}_\theta$, Eq.~\eqref{eq::sysEvo}.
However, this introduces an arbitrariness on the choice of initial state $\rho_S$, which can in principle be prepared in any way. This leads to a clear problem in comparing with the sequential setup since the states in that case are only prepared once and subsequently evolve. The second is to obtain iid outcomes by resetting the state of the probe system after each measurement. This could mean, for example, coupling it to a heat bath after each measurement, hence erasing information about past outcomes. Once again, this introduces an arbitrariness as to how the reset occurs and an additional arbitrary parameter, which is  the time it takes to re-prepare the system. 

In Ref.~\cite{De_Pasquale_2017}, the authors compared their results with the iid scenario. In particular, they considered the situation in which the system was always re-prepared in specific states $|k'\rangle$. 
The corresponding Fisher information is then a single term in the sum appearing in Eq.~\eqref{eq:seqFI2},
\begin{equation}\label{conditional_Fisher_projective}
    F_{2|1=k'} := \sum_{k} \frac{1}{P(k|k')} \left(\frac{\partial}{\partial \theta} P(k|k')\right)^2.
\end{equation}
It is clear from Eq.~\eqref{eq::F21} that $F_{2|1}$ will be a convex sum of such quantities:
\begin{equation}
    F_{2|1} = \sum_{k'} q_{k'} F_{2|1=k'}.
\end{equation}
We can therefore have $F_{2|1=k'}\! \lessgtr \! F_{2|1}$, depending on the particular choice of $k'$. The quantities $F_{2|1=k'}$ are useful, as they tell us which outcomes $k'$ lead to higher information gains. 
However, if one is using just a single probe then the quantity in fact being sampled is $F_{2|1}$. 

\subsection{Comparison with $F_1$}
Alternatively, we may be inclined to compare $F_{2|1}$ with $F_1$. The former is the information contained in the transition probabilities $P(k|k')$ and the latter is the information contained in distribution $q_k$ [Eq.~\eqref{Markov_chain_steady_state_equation}]. However, this comparison is generally not fair since the $q_k$-information is not acquired over independent trials. Instead, it is determined sequentially in a single run. This subtle point was recently discussed by some of us in Ref.~\cite{Smiga2023} and can be clarified as follows. The actual data we have at hand is the string $k_1,\ldots,k_N$. Estimation therefore proceeds by building a function $\hat{\theta}(k_1,\ldots,k_N)$ to use as the estimator. For any (unbiased) estimator, the error for large $N$ will be bounded by $1/(N F_{2|1})$. 

To achieve this bound, however, we must use estimators that make use of the transitions. For example, suppose that the functional form of a specific transition reads $P(2|1)\!=\! f(\theta)$, for some function $f(\theta)$. Then a potential estimator could be constructed as follows: given a single string $k_1,\ldots,k_N$, we count how many times $k\!=\!1$ was followed by $k\!=\!2$, and use this to build an estimate $\hat{P}(2|1)$ for the transition probability. The function $f^{-1}(\hat{P}(2|1))$ would then be an estimator of $\theta$, which will generally be unbiased for large $N$. Since this estimator uses information about transitions, it {\it might} saturate the Cram\'er-Rao bound asymptotically, although there is no guarantee of this. 

In practice, we might prefer to use simpler estimators. For example, we can build a histogram of the outcomes. That is, given a string $k_1,\ldots,k_N$, where each $k_i$ ranges over some alphabet $k_i\in \{1,\ldots, d\}$, we can build a histogram counting how meany times $k_i = 1$ is recorded, how many times $k_i = 2$ is recorded and so on. This is called the {\it empirical distribution} (ED) and is a form of data compression. It can be shown that the ED is an unbiased estimator of the steady-state probabilities $q_k$, and hence, naively, we might expect that the information associated to it should be $F_1$. However this is not the case, as shown in~\cite{Smiga2023}. The reason being that the data string $k_1,\ldots,k_N$ is {\it not} iid. Instead, it is acquired sequentially on a single run. As a consequence, due to correlations between sequential outcomes, the resulting Fisher information is affected. The actual form for the Fisher information in the ED is described in Ref.~\cite{Smiga2023}.

The only way we would obtain an information rate given by $F_1$ is if we perform the same protocol as Sec.~\ref{ssec:projective}, but only use data points spaced by a large distance $\Delta \gg 1$. 
That is, we would have to perform $N\Delta$ measurements, however, instead of building an estimator based on $k_1,\ldots,k_{N\Delta}$, we discard intermediate data and build an estimator involving only $k_1, k_{\Delta+1},k_{2\Delta+1},\ldots$. This, of course, is a terrible strategy since it involves throwing away valuable data. 

\subsection{Direct measurements on the environment}
In our approach, information about a parameter of the environment is obtained by coupling it to a probe system via the map~\eqref{eq::sysEvo}. Suppose that the only dependence on $\theta$ is in the environment's state $\rho_E^\theta$. Then Eq.~\eqref{eq::sysEvo} represents a form of data compression; that is, information is lost when it is transferred from $\rho_E^\theta$ to $\rho_S$. A natural way to quantify the amount lost is to compute the quantum Fisher information of $\rho_E$, which already maximizes over all possible measurements on the environment. The resulting quantity must then necessarily exceed $F_{2|1}$. In reality, of course, this compression can be significant, for example if the environment is very large and the system is small. In Appendix~\ref{AppCG} we prove a stronger result: we consider a maximization only over measurements that have the same number of outcomes as the dimension of the system. 
We find that
\begin{equation}
    F_{2|1} \leq F_\theta^{\rm iid^*} \leq F(\rho_E,G_i^*),
    \label{eq::CGFisher}
\end{equation}
where $F_\theta^{\rm iid^*}$ is Eq.~\eqref{fisherinfo1} maximised over all initial states and $\{G_i^*\}$ is the optimal POVM with the same number of measurement outcomes as the dimension of $\rho_S$. Hence, even restricting the number of outcomes in the environment, a direct measurement would still be better than using a probe. This agrees with the results of Ref.~\cite{Hovhannisyan_2021}, which studied temperature estimation in thermal states. 

\subsection{Relation to collisional schemes}
The sequential measurement approach shares several commonalities with the recently proposed framework of collisional thermometry~\cite{Seah2019,Shu2020,Alves_2022,O_Connor_2021}. Instead of performing the measurement on the system itself, an auxiliary system is used which interacts (collides) with the probe system and the measurement is subsequently made on this auxiliary system. For projective measurements performed immediately after the interaction, we find that the Fisher information is equivalent to the sequential measurement scheme. We assume that the system and colliding auxiliary unit are initially uncorrelated and evolve via a unitary interaction
\begin{equation}
    \rho_{SC} = U(\rho_S \otimes \rho_C)U^\dag,
\end{equation}
after which we perform a measurement of the auxiliary unit in an arbitrary basis $\{\ket{i}\}$. The probability of getting a measurement result $i$ is then given by
\begin{align}
    \text{P}(i) &= \Tr[\rho_{SC}(\mathbf{I}_S\otimes \ketbra{i})] \nonumber\\ \nonumber
    &=\Tr_S\left[\sum_j F_{i,j} \rho_S F_{i,j}^\dag\right]\\ 
    &=\Tr_S\left[\rho_S E_i\right],
\end{align}
where we have defined $F_{i,j} \!=\! \mel{i}{U}{p_j}$ with $\ket{p_j}$ an eigenvector of the auxiliary unit and $E_i \!=\! \sum_j F_{i,j}^\dag F_{i,j}$ . Through a similar analysis to the one in Appendix~\ref{AppCG} we can prove that $\{E_i\}$ is a POVM on $\rho_S$. Therefore, performing a measurement on the auxiliary unit immediately after interaction is equivalent to performing a (different) measurement on the system itself. In fact, Neumark's theorem~\cite{gelfand1943imbedding} proves that any POVM on the system can be realised via a suitable projective measurement on a collisional unit. This may be a useful practical method of realising some more complicated forms of POVMs on the system. The collisional setup still provides some additional freedom to make use of initial correlations between measurements~\cite{Shu2020} or collective measurements on multiple collisional units but a significant advantage has yet to be demonstrated for these methods.

\section{Applications}
\label{ApplicationsSec}
\subsection{Precision Thermometry}
\label{sec:thermometryI}
We now turn to applications of our formalism. First, we consider the case of quantum thermometry~\cite{De_Pasquale_2018,Razavian_2019,Mehboudi_2019,Mitchison_2020,Brattegard2023,GeorgePRA,Mok_2021}, where it is known that an advantage can be obtained by using quantum probes for low temperatures~\cite{Correa_2017,Hovhannisyan_2018,Potts_2019}. Nevertheless, estimation of thermal probes is limited by the thermal Fisher information~\cite{Correa_2015} which is maximised by using a $D$ level probe with a non-degenerate ground state and $(D-1)$-degenerate excited states~\cite{Correa_2015}. The Hamiltonian reads $H_p = e_0\ketbra{e_0} + \sum_{i=1}^{D-1}e_1\ketbra{e_i}$, with energy spacing $e_1-e_0\!=\! \Omega$. Following Ref.~\cite{Correa_2015}, we model the environment as a bosonic heat bath with a flat spectral density. In suitable limits, this leads to the following master equation
\begin{equation}
    \frac{d\rho_S}{dt} = \mathbb{L}\rho_S =  \gamma \sum_{i=1}^{D-1} \Big\{ (1+\bar{n})
    \mathcal{D}\big[ |e_0\rangle\langle e_i| \big]
    + \bar{n} \mathcal{D}\big[ |e_i \rangle\langle e_0 \big]\Big\}\rho_S,
    \label{eq::envmaster}
\end{equation}
where $\mathcal{D}[L] \rho = L \rho L^\dagger - \frac{1}{2}\{L^\dagger L,\rho\}$,  $\gamma$ is the system-environment coupling and $\bar{n} = 1/(e^{\hbar \Omega / k_B T} -1)$ is the mean occupation number. 
Our goal will be to estimate the occupation $\bar{n}$, from which we can estimate $T$ assuming $\Omega$ is fixed and known~\cite{GeorgeMPEst}. The map $\mathcal{E}$, Eq.~\eqref{eq::sysEvo}, corresponds to the evolution $\rho_S(t) = \mathcal{E}(\rho_S(0)) = e^{\mathbb{L}\tau} \rho_S(0)$, up to a certain time $\tau$. This defines the time between measurements and will be used as a free parameter of the model which can be optimized over to enhance the estimation precision. We will restrict to measurements in the energy basis, since this is known to be optimal in the case of incoherent states. As a consequence, the system remains diagonal throughout the protocol. 

In the limit of infinite evolution time $\tau\!\to\!\infty$, the steady state of the master equation~\eqref{eq::envmaster} is the Gibbs thermal state, with ground-state occupation $q_0 = \langle e_0 |\rho_S^{\rm ss}| e_0\rangle = \frac{1+\bar{n}}{1+ D \bar{n}}$, and excited state occupation $q_i = \langle e_i | \rho_S^{\rm ss} |e_i\rangle =  \frac{\bar{n}}{1+ D \bar{n}}$, for all states $i = 1,\ldots,D-1$. In this limit the map $\mathcal{E}$ completely resets the state of the system, causing the outcomes to become iid. The corresponding Fisher information, which is the analog of Eq.~\eqref{fisherinfo1}, is given by
\begin{align}
 \label{thermalFI}
    \mathcal{F}_{th} &= \frac{1}{q_0} \left(\frac{\partial}{\partial \bar{n}}q_0\right)^2+(N-1)\frac{1}{q_1} \left(\frac{\partial}{\partial \bar{n}}q_1\right)^2 \\ \nonumber
    &= \frac{D-1}{\bar{n}(1+\bar{n})(1+D \bar{n})^2},
\end{align}
and is precisely the thermal Fisher information. Particular to this setup, Eq.~\eqref{thermalFI} coincides with all of the comparison quantities in Sec.~\ref{sec:comparisons}, making it the logical benchmark for the correlated process.

Similar in spirit to the collisional thermometry scheme of Ref.~\cite{Seah2019}, with sequential measurements we can exploit the additional information about temperature that is contained in the thermalization rates of the probe by relaxing the assumption that $\tau\!\!\to\!\!\infty$ between each measurement, therefore the probe only partially thermalizes after each measurement. The first step is  to compute the transition probability $P(k|k')$ in Eq.~\eqref{transition_probabilities}, where $|k\rangle$ now corresponds to the energy basis $|e_k\rangle$ of the probe Hamiltonian. These probabilities can be explicitly calculated, as detailed in Appendix~\ref{App:Thermometry}, and we find 
\begin{align}\label{thermometry_P00}
    P(0|0) &= 1-q_e(1-f)
    \\[0.2cm]
    P(i|0) &= \frac{q_e(1-f)}{D-1}
    \\[0.2cm]
    P(0|i) &= q_0 (1-f) 
    \\[0.2cm]
    P(i|j) &= \frac{q_e(1-f) + f - g}{D-1},\qquad j\neq i
    \\[0.2cm]
    P(i|i) &= g + P(i|j),
\label{thermometry_Pii}
\end{align}
where we have defined the probability of finding the system in any of the excited states, $q_e =1-q_0$,
as well as the functions $g = e^{-\gamma \tau(\bar{n}+1)}$ and $f = e^{-\gamma \tau(D\bar{n}+1)}$ which represent the two most relevant relaxation rates of the problem. We remark that these rates naturally depend both on the choice of $\tau$, as well as on $\bar{n}$, but we omit these explicit dependences for clarity of notation. The steady-state of the Markov chain, Eq.~\eqref{Markov_chain_steady_state_equation} is given by the same equilibrium probabilities $q_k$ defined above. This is not immediately obvious: while it certainly must hold true when $\tau\to\infty$, for finite times this is less evident. The rationale behind this is explained in detail in Appendix~\ref{App:Thermometry}. 

Using these results we can compute the conditional Fisher information rates $F_{2|1=k'}$ in Eq.~\eqref{conditional_Fisher_projective}. If the measurement outcome was $k'\!=\!0$ (i.e., the system was found in the ground-state), then the Fisher information rate for the next measurement will be 
\begin{align}
    F_{2|1=e_0} &= \frac{[\partial_{\bar{n}} P(0|0)]^2}{P(0|0)}
    + (D-1) \frac{[\partial_{\bar{n}} P(0|1)]^2}{P(0|1)}
    \\[0.2cm]
    &= \frac{(\partial_{\bar{n}} x)^2}{x(1-x)},
    \label{thermometry_F21e0}
\end{align}
where $x = q_e(1-f) = 1-P(0|0)$ is the probability that, after observing the system in the ground state, it is excited to any of the excited states after a time $\tau$. This conditional Fisher information rate is therefore the same as that of a binary random variable, where the system is either in the ground or in the excited state after a time $\tau$, given that at time $t\!=\!0$ it was in the ground state. Similarly, we can calculate the conditional Fisher information rate given that the outcome was one of the excited states $i=1,\ldots,D-1$. From Eq.~\eqref{conditional_Fisher_projective} it follows that 
\begin{equation}
    F_{2|1=e_i} = \frac{[\partial_{\bar{n}} P(0|1)]^2}{P(0|1)} + 
    \frac{[\partial_{\bar{n}} P(1|1)]^2}{P(1|1)}
    +(D-2)\frac{[\partial_{\bar{n}} P(2|1)]^2}{P(2|1)},
\end{equation}
where the factor of $D-2$ represents the number of excited states the system can go to, other than $i$, given that it was initially detected in $i$. With some simplifications, the final expression reads 
\begin{align}
   \nonumber
   F_{2|1=e_i} = \frac{(\partial_{\bar{n}} y)^2}{y} 
    &+ \frac{1}{D-1} \frac{[(\partial_{\bar{n}} y) - (D-2) (\partial_{\bar{n}} g)]^2}{1- y - (D-2) g}
    \\[0.2cm]
&+ \frac{D-2}{D-1} \frac{[(\partial_{\bar{n}}y + \partial_{\bar{n}} g]^2}{1-y - g},
    \label{thermometry_F21ei}
\end{align}
where $y = 1- x - f$. The first term is the Fisher information of the binary process $i\to 0$, the second is the rate for $i\to i$, and the third is the rate for $i\to j$ with $j\neq i$, weighted by the $D-2$ possible $j$'s. 

\begin{figure}[t]
\hskip0.048\columnwidth(a) \hskip0.417\columnwidth (b)\\
\includegraphics[width=0.45\columnwidth]{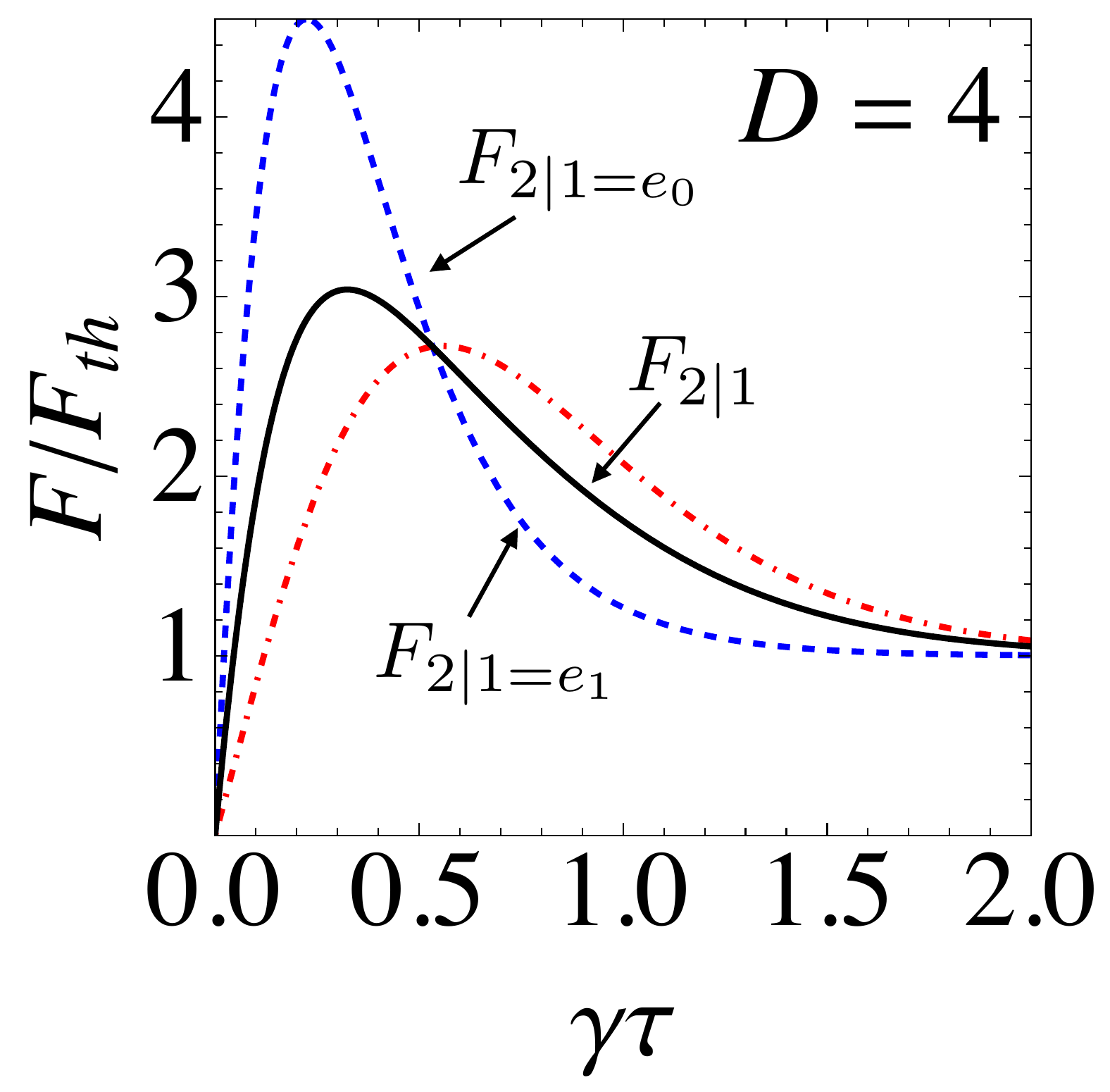}~~\includegraphics[width=0.45\columnwidth]{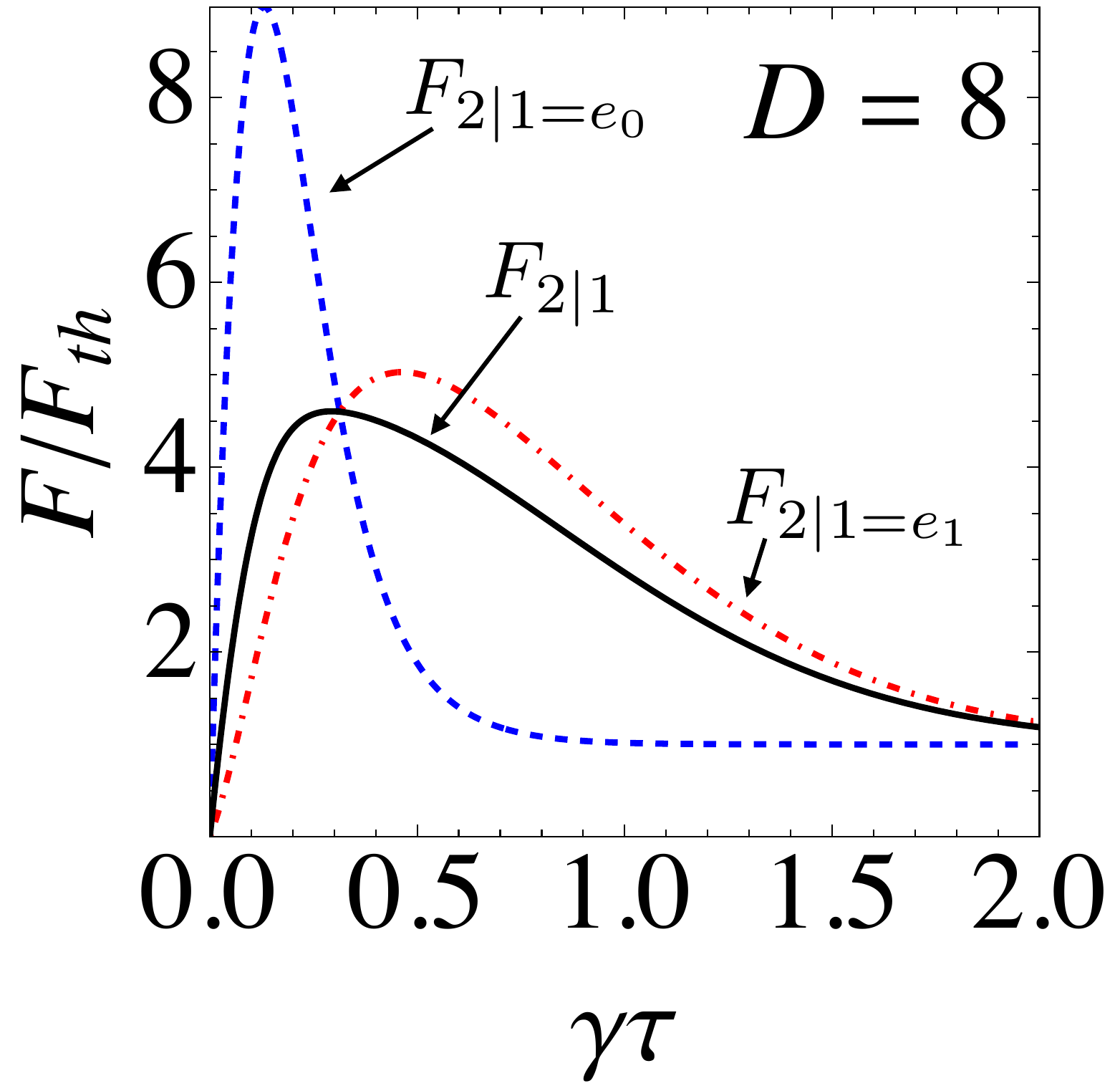}\\
(c)\\
\includegraphics[width=0.9\columnwidth]{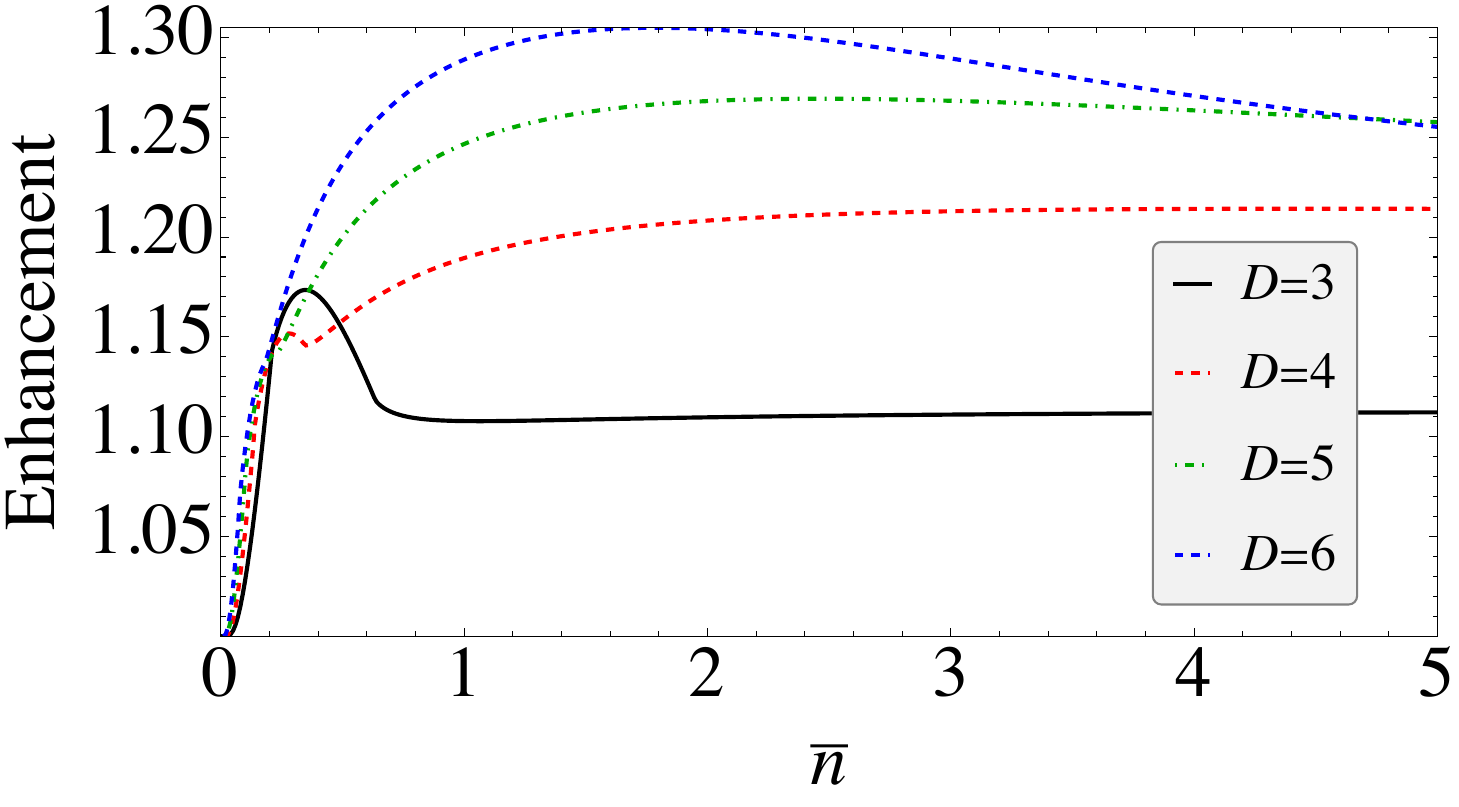}
\caption{
Ratio of the relevant Fisher informations from the sequential scheme to the thermal Fisher information. We show $F_{2|1}$ [solid] and the individual contributions to this Fisher information for specific previous measurement results $F_{2|1=e_0}$ [dashed], and $F_{2|1=e_1}$ [dot-dashed] of the $D-1$ degenerate probe as a function of the time between measurements for arbitrary choice of $\bar{n} = 1$ and {(a)} $D=4$ and (b) $D=8$. (c) Maximum enhancement, $F_{2|1}^\# /F_{2|1}^*$, for the optimised protocol leveraging information about the previous measurement outcome} as a function of $\bar{n}$. We show the results for different values of $D=3$ [bottom-most, solid], 4 [dotted], 5 [dot-dashed], and 6 [top-most, dashed].
\label{fig:F21}
\end{figure}

From these results, the total Fisher information rate in Eq.~\eqref{eq:seqFI2} is 
\begin{equation}
    F_{2|1} = q_0 F_{2|1=e_0} + q_e F_{2|1=e_i},
    \label{thermometry_F21}
\end{equation}
where, recall, $q_e = (D-1) q_i$. The quantities $F_{2|1=e_0}$, $F_{2|1=e_i}$ and $F_{2|1}$ are shown in 
Fig.~\ref{fig:F21} (a) and (b) as a function of the measurement time $\gamma\tau$. We analyze the three different contributions, Eqs.~\eqref{thermometry_F21e0}, \eqref{thermometry_F21ei} and~\eqref{thermometry_F21} for $D=4$ and $D=8$. It is clear that there is an optimal finite time at which the Fisher information rates are maximized, i.e. partial thermalization is favorable as it allows us to gain more information about the temperature of the system from the relaxation rates. We can find this optimal Fisher information by calculating $F_{2|1}^* = \max_\tau F_{2|1}$ for any given value of $\bar{n}$.

From Fig.~\ref{fig:F21}(a) it is clear that the value of the inter-measurement times $\gamma \tau$ which give the highest precision depend on whether the outcome was the ground or the excited state. Based on this, we can therefore envision a feedback mechanism that chooses the value of $\tau$ depending on the outcome. This would amount to using different values of $\tau$ in the probabilities~\eqref{thermometry_P00}-\eqref{thermometry_Pii}. However, some care must be taken in doing so because the steady-state distribution $q_k$ will no longer be the thermal distribution. As a consequence, the optimal times are not exactly the peaks of the dashed lines in Fig.~\ref{fig:F21}(a) and (b). The new steady state probabilities are given by
\begin{equation}
    q_0 = \frac{(1-f(\tau_g)) \bar{n}}{1+D \bar{n}(1-f(\tau_g)) + f(\tau_g) \bar{n} - (1 + \bar{n})f(\tau_e)}.
\end{equation}
To determine the maximal achievable precision we must optimise the combined Fisher information, Eq~\eqref{thermometry_F21}, over both $\tau_e$ and $\tau_g$, giving $F_{2|1}^\# = \max_{\tau_g,\tau_e} F_{2|1}$, which is a complex optimization problem which must be solved numerically. In Fig.~\ref{fig:F21}(c) we show the enhancement achievable, i.e. $F_{2|1}^\# /F_{2|1}^*$, as a function of $\bar{n}$, which also scales with the increasing dimension of the probe. We find that the ratio of the optimal $\tau_g$ and $\tau_e$ changes very little with temperature implying that the feedback mechanism can reliably achieve this enhancement.

\begin{figure}[t]
\hskip0.048\columnwidth(a) \hskip0.417\columnwidth (b)\\
\includegraphics[width=0.45\columnwidth]{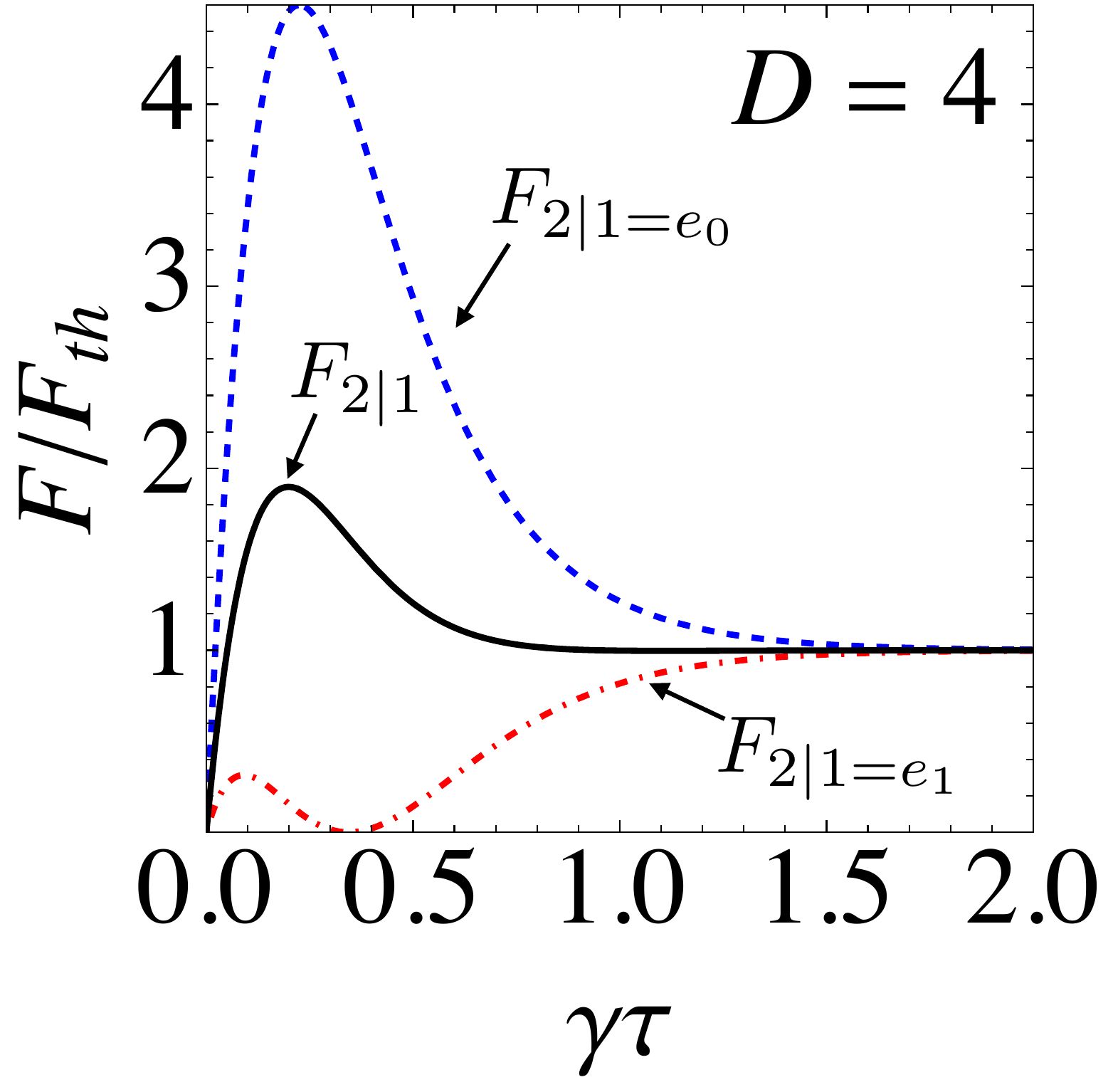}~~\includegraphics[width=0.45\columnwidth]{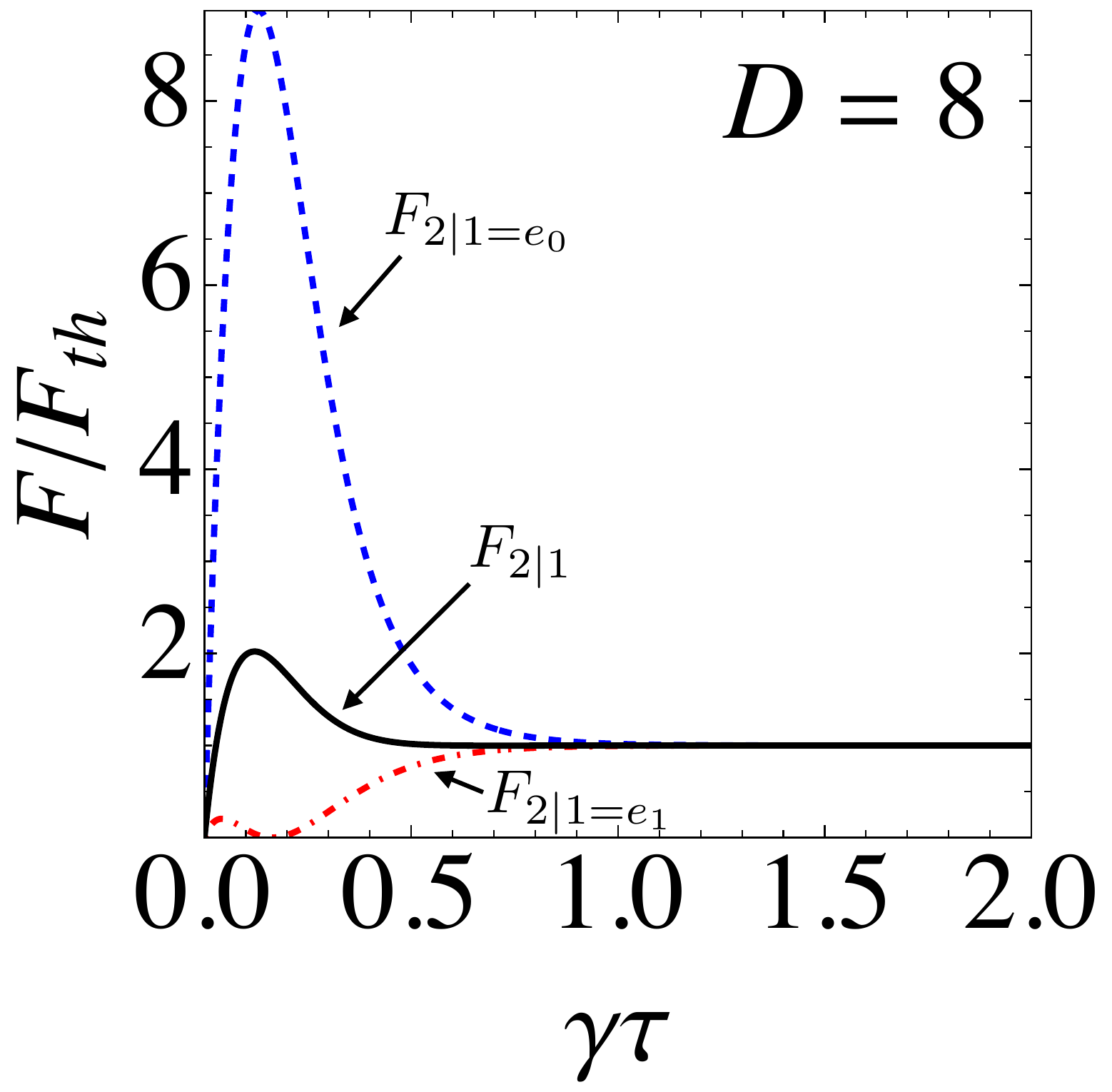}\\
(c)\\
\includegraphics[width=0.9\columnwidth]{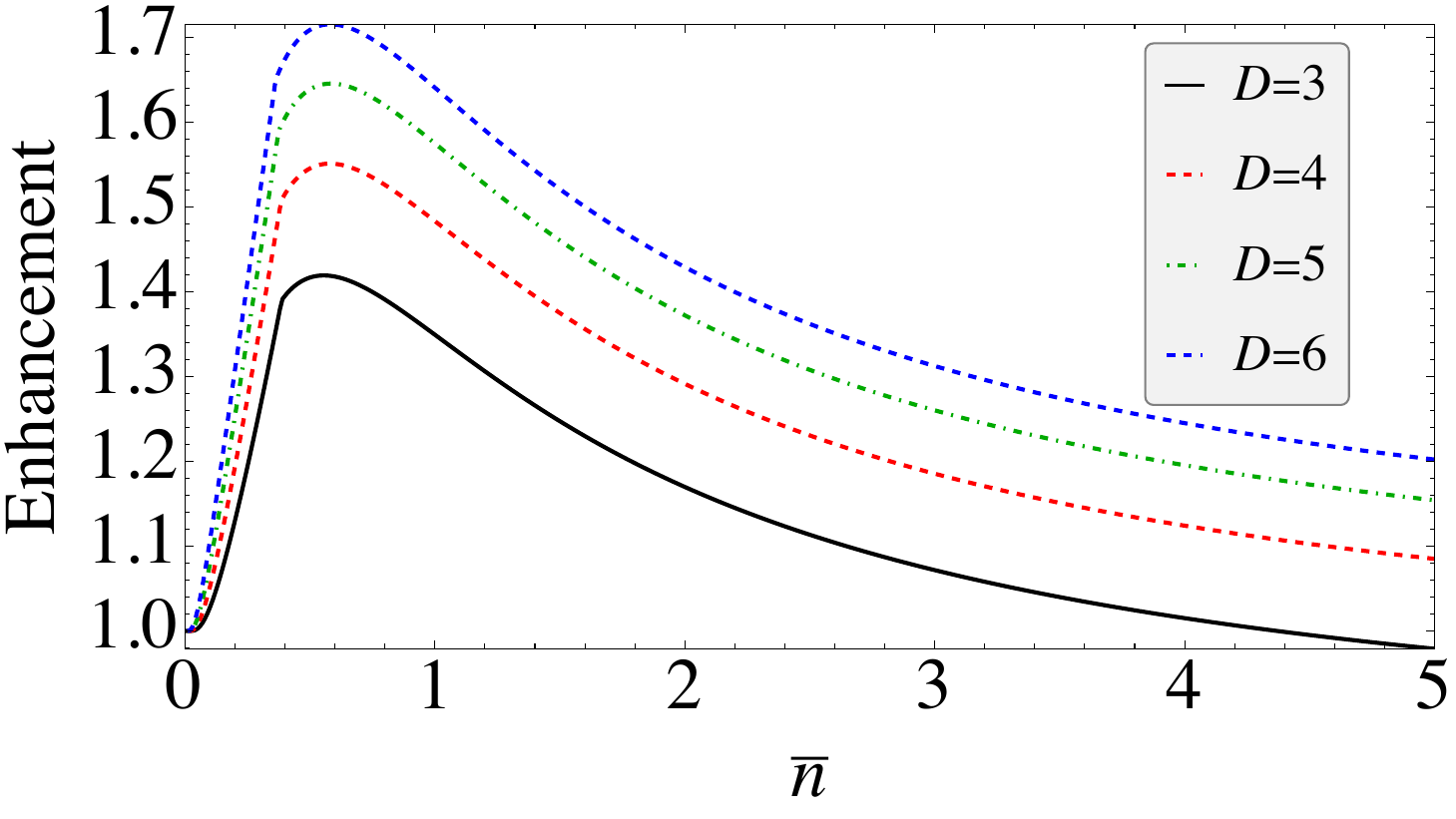}
\caption{
As for Fig.~\ref{fig:F21}, however, considering coarse grained energy measurements which cannot distinguish between measurement outcomes in the degenerate eigenspace.}
\label{fig:F21Energy}
\end{figure}

\subsection{Themometry with Coarse Grained Energy Measurements}
\label{sec:thermometryII}
Due to the degeneracy of the probe, there is a subtle distinction when we only have access to an energy measurement which is unable to distinguish between measurement outcomes in the degenerate eigenspace and which we will call partially indiscriminate, instead of the full measurement in the energy basis considered in Sec.~\ref{sec:thermometryI}. The result of such a 
partially indiscriminate energy measurement would be the POVM
\begin{equation}
    E_0 = \ketbra{e_0};\quad E_1 = \sum_{i=1}^{D-1}\ketbra{e_i}.
\end{equation}
Although this is no longer a projective measurement, the Fisher information still retains the same form as a projective measurement, i.e. that of Eq.~\eqref{eq::F21}. This is because the transition probability from one energy eigenspace to another is independent of the the specific state that the system is initially in within that subspace. The transition probabilities can be calculated in a analogous manner to the full energy basis measurement case detailed before and we find
\begin{align}\label{thermometry_P00_Energy}
    P(0|0) &= 1-q_e(1-f)
    \\[0.2cm]
    P(1|0) &= q_e(1-f)
    \\[0.2cm]
    P(0|1) &= q_0 (1-f) 
    \\[0.2cm]
    P(1|1) &= 1- q_0 (1-f).
\end{align}
As anticipated, the transition probabilities here are identical for transitions to the ground state and the transition probabilities to the excited subspace are simply the sum of all the transition probabilities to the individual excited energy eigenstates. The main difference now is that we are unable to distinguish measuring the {\it same} excited energy eigenstate two measurements in a row from measuring two different excited energy eigenstates. We can see the consequences of this difference in Fig~\ref{fig:F21Energy}(a) and comparing it with Fig.~\ref{fig:F21}(a). We see that the Fisher information is exactly the same when the previous measurement was a ground state (upper blue curve) but it is significantly lower for previous measurements that resulted in an excited state outcome due to the reduction in information we have for this measurement 
(bottom-most red dot-dashed curve). For low values of $\bar{n}$ the Fisher information attainable when the previous measurement was excited is never larger than the thermal Fisher information. In this case an obvious measurement strategy to pursue would be to allow the system to fully thermalise after an excited energy measurement outcome is recorded and then optimise the Fisher information over the measurement time after a ground state measurement outcome is obtained. In fact, this is the optimal strategy for small $\bar{n}$ and we can see in Fig~\ref{fig:F21Energy}(c) that it can still provide a significant advantage over any strategy without feedback control.

This strategy indicates that there are other possible metrology protocols that, while not being projective measurements, nonetheless maintain the same form of the Fisher information as seen in Eq.~\eqref{eq::F21}. For Eq.~\eqref{eq::F21} to hold, outcomes must depend only on the result of the directly preceding measurement. While projective measurements are an example of such a process, they are not the only example. As just discussed, we satisfy this condition if all of the POVM operators project onto degenerate subspaces. Additionally if we have POVM operators that project onto a non-degenerate subspace we could allow the system to fully equilibrate after obtaining that measurement result, the system will therefore be in the equilibrium state, independent of any previous measurements.

\subsection{Rabi Frequency Estimation}
We next demonstrate that the sequential metrology approach can be employed to determine a property of the system Hamiltonian, such as the Rabi frequency of a driven qubit~\cite{Kiilerich_2015}, thus extending its applicability beyond estimating parameters of only the environment. To make our ideas concrete, we consider a qubit probe with Hamiltonian $H_S \!=\! \Omega \sigma_x$ which is coupled to an environment according to the master equation 
\begin{equation}\label{QME_Rabi}
    \frac{d\rho_S}{dt} =\mathbb{L} \rho_S = -i [H,\rho_S] + \gamma \mathcal{D}[\sigma_-]\rho_S.
\end{equation}
We once again consider sequential projective measurements in the system, with a free evolution of duration $\tau$ in between. Equation~\eqref{transition_probabilities} then becomes 
\begin{equation}
    P(k|k') = \langle k |\Big( e^{\mathbb{L}\tau} (|k'\rangle\langle k'| \Big)|k\rangle. 
\end{equation}
From this we can determine the steady-state distribution $q_k$ in Eq.~\eqref{Markov_chain_steady_state_equation} and subsequently determine the conditional Fisher information rates $F_{2|1=k'}$ in  Eq.~\eqref{conditional_Fisher_projective}, as well as its average $F_{2|1}$ in Eq.~\eqref{eq::F21}.

\begin{figure}
    \centering
    (a)\\
    \includegraphics[width=0.9\columnwidth]{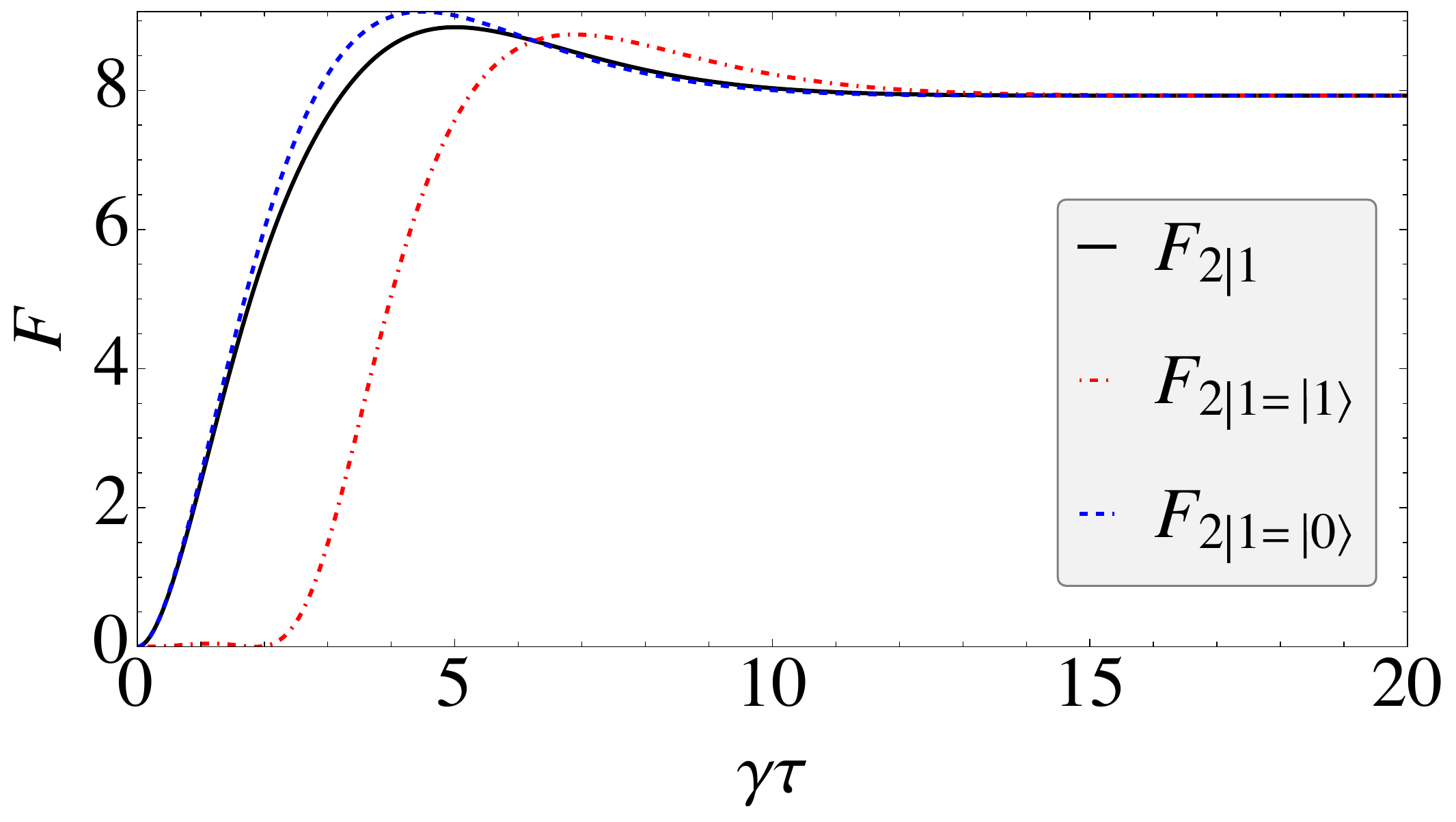}\\
    (b)\\
    \includegraphics[width=0.9\columnwidth]{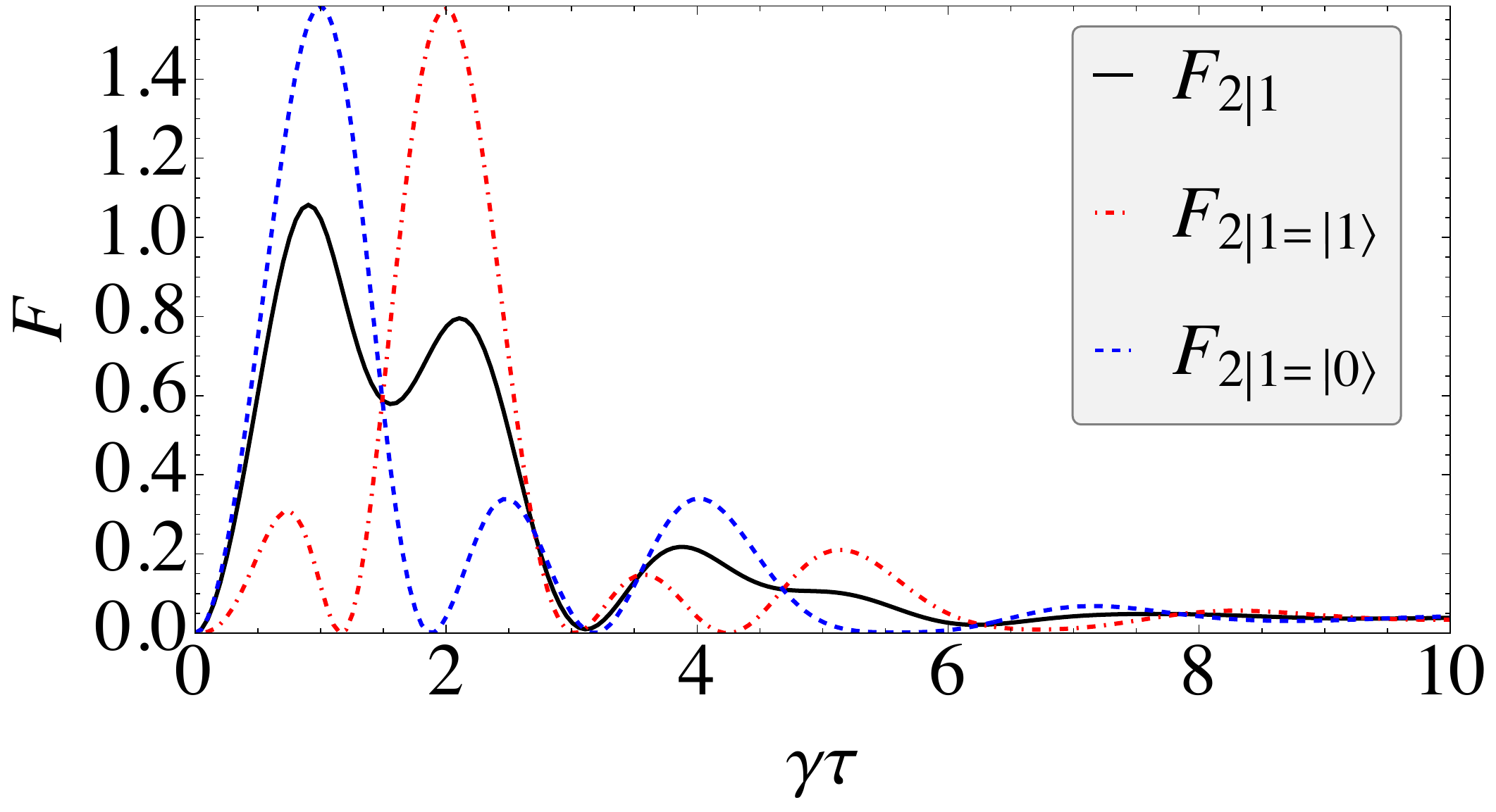}
\caption{Fisher information from the sequential scheme for estimating the Rabi frequency $\Omega$ in the Rabi model~\eqref{QME_Rabi}. We show $F_{2|1}$ [solid] and the individual contributions to this Fisher information for specific previous measurement results $F_{2|1=\ket{0}}$ [dashed], and $F_{2|1=\ket{1}}$ [dot-dashed] for (a) $\Omega/\gamma = 0.2$ and (b) $\Omega/\gamma = 1.0$. The measurements are performed in the computational basis.}
    \label{fig:rabi_plot1}
\end{figure}

Fig.~\ref{fig:rabi_plot1} shows results for $F_{2|1=k'}$ and $F_{2|1}$ in the case of measurements in the computational basis $k'=\{\ket{0} , \ket{1} \}$ and we show the results for $\Omega/\gamma\!=\! 0.2$ in panel (a) and $\Omega/\gamma = 1.0$ in (b). The eigenvalues of the Liouvillian depend on $\sqrt{\gamma^2 - 64 \Omega^2}$, which becomes imaginary for $\Omega/\gamma\!>\! 1/8$. This is evident by comparing the behavior between the two settings in Fig.~\ref{fig:rabi_plot1} as all Fisher rates show significantly more oscillations for larger $\Omega$. We also see from the images that the information rates depend sensibly on $\tau$, and this becomes particularly strong for large $\Omega/\gamma$, cf. Fig.~\ref{fig:rabi_plot1}(b), to the point where the Fisher information can actually be zero at certain points.

One could also study the same problem for bases in other directions. Measurements in the $\sigma_x$ basis yield no information, while measurements in $\sigma_y$ can and, in fact, generally do lead to somewhat larger Fisher information rates, although their behavior with $\gamma\tau$ is also different. Finally, one could ask about what is the optimal basis. However, this quickly becomes a difficult problem to solve in general since the basis will depend on the actual value of $\Omega$, as well as on $\tau$.

\section{Conclusions}
We have examined how temporal correlations established between measurement outcomes impact the achievable precision in estimating a parameter of interest using quantum probes. We considered a sequential measurement protocol, where the probe system is stroboscopically measured. We established that the resulting conditional Fisher information captures the rate at which information about the parameter of interest can be obtained. For protocols employing projective measurements, we have used our formalism to demonstrate that advantageous schemes can be developed. In the case of thermometry we showed that allowing for different waiting times between measurements of the probe based on the previous measurement outcomes allows for an increase in the achievable precision. Furthermore, we demonstrated that the protocol is versatile, allowing to effectively estimate Hamiltonian parameters such as the Rabi frequency. The latter example also established that the choice of measurement can play a significant role in the achievable precision, thus opening the possibility to explore whether further enhancement can be achieved by extending an adaptive scheme beyond allowing for different measurement times but also implementing measurements in a different basis at each step.

This work builds on Refs.~\cite{Radaelli_2023, Smiga2023}, which studied stochastic metrology of generic correlated outcomes in classical processes and connects those results with quantum processes. In particular, with the stochastic outcomes obtained when a quantum system is subject to stroboscopic measurements. Our work highlights the subtlety and care that must be considered when measurement outcomes in a metrological protocol are not independent and identically distributed. Furthermore, it provides a useful framework to explore a wider class of sensing protocols, in particular those that can leverage the temporal correlations to be more metrologically effective.

\acknowledgements
This work was supported by the Science Foundation Ireland Starting Investigator Research Grant ``SpeedDemon" No. 18/SIRG/5508 and the John Templeton Foundation Grant ID 62422. SC is grateful to the Alexander von Humboldt Foundation for their support.

\appendix
\setcounter{equation}{0}
\renewcommand\theequation{A.\arabic{equation}}
\section{Coarse grained comparison}
\label{AppCG}

We consider the setup from Eq~\eqref{eq::sysEvo} of a general system-environment evolution with $U$ and $\rho_S$ independent of $\theta$ and $\rho_S = \sum_i s_i\ketbra{s_i}$. When this is not the case it is possible do violate the following bound. The addition of an auxiliary system such as was considered in Ref.~\cite{Hovhannisyan_2021} has no effect on the following proof as long as the auxiliary system also has no $\theta$ dependence. Our first step is to derive the map $\mathcal{E}'(\rho_E) = \mathcal{E}(\rho_S)$.

\begin{align}
\mathcal{E}(\rho_S) &= Tr_E[U(\rho_S \otimes \rho_E)U^\dag] \\ \nonumber
&= \sum_i \tensor[_E]{\bra{b_i}}{}U\left(\sum_j s_j\tensor[_S]{\ketbra{s_j}}{_S} \otimes \rho_E\right)U^\dag\tensor[]{\ket{b_i}}{_E}\\ \nonumber
&= \sum_{i,j} s_j \tensor[_E]{\mel{b_i}{U}{s_j}}{_S} \rho_E \tensor[_S]{\mel{s_j}{U^\dag}{b_i}}{_E}\\ \nonumber
&= \sum_{i,j} M_{i,j}\rho_E M_{i,j}^\dag \\ \nonumber
&= \mathcal{E}'(\rho_E),
\end{align}
where $\{\ket{b_i}\}$ is an arbitrary basis of $\mathcal{H}_E$ and $M_{i,j} = \sqrt{s_j}\tensor[_E]{\mel{b_i}{U}{s_j}}{_S}$. For the following analysis we require that $M_{i,j}$ is independent of $\theta$ with is clearly true when $U$ and $\rho_S$ are independent of $\theta$ but can also be true even when this is not the case. The quantum Fisher information of $\mathcal{E}(\rho_S)$ is then given by the Fisher information of $p_i(\mathcal{E}(\rho_S),E_i) = \Tr[\mathcal{E}(\rho_S) E_i]$ maximised over all possible POVMs $\{E_i\}$. Where $\{E_i\}$ is a set of hermitian, positive semi-definite matrices that sum to the identity. Lets now look at the quantum Fisher information of $\mathcal{E}(\rho_S) = \mathcal{E}'(\rho_E) = \sum_j M_j\rho_E M_j^\dag$. We will label the optimal POVM as $\{F_i\}$.
\begin{align}
p_i(\mathcal{E}(\rho_S),F_i) &= p_i(\mathcal{E}'(\rho_E),F_i) \\ \nonumber
&= \Tr[\mathcal{E}'(\rho_E) F_i] = \Tr[\sum_{j,k} M_{j,k}\,\rho_E M_{j,k}^\dag F_i]\\ \nonumber
&= \sum_{j,k} \Tr[M_{j,k}\,\rho_E M_{j,k}^\dag F_i] = \sum_j \Tr[\rho_E M_{j,k}^\dag F_i M_{j,k}]\\ \nonumber
&= \Tr\left[\rho_E\left(\sum_{j,k} M_{j,k}^\dag F_i M_{j,k}\right)\right].
\end{align}
We can now define a new set of operators $G_i = \sum_{j,k} M_{j,k}^\dag F_i M_{j,k}$, it is important to note that $\{G_i\}$ has the same number of elements as  $\{F_i\}$. Now we need to prove that $\{G_i\}$ is a valid POVM on $\mathcal{H}_E$ . Since $F_i$ is Hermitian then $G_i$ clearly is too. A matrix is positive semi-definite if and only if it can be decomposed into a product $F_i = L_i^\dag L_i$. Since $\{F_i\}$ is a POVM we know that it can be decomposed. Therefore we can write $G_i = \sum_{j,k} M_{j,k}^\dag L_i^\dag L_i M_{j,k} = \sum_{j,k} K_{i,j,k}^\dag K_{i,j,k}$ with $K_{i,j,k} = L_iM_{j,k}$. This means $G_i$ is the sum of positive semi-definite matrices and is therefore also positive semi-definite. The last thing to show is that $\{G_i\}$ sums to the identity
\begin{align}
\sum_i G_i &= \sum_i\sum_{j,k} M_{j,k}^\dag F_i M_{j,k}\\\nonumber
&= \sum_{j,k} M_{j,k}^\dag \left(\sum_i F_i\right) M_{j,k} = \sum_{j,k} M_{j,k}^\dag M_{j,k}\\ \nonumber
&= \sum_{j,k} s_k \tensor[_S]{\mel{s_k}{U^\dag}{b_j}}{_E} \tensor[_E]{\mel{b_j}{U}{s_k}}{_S}\\ \nonumber
&= \sum_{k} s_k \tensor[_S]{\bra{s_k}}{} U U^\dag \tensor[]{\ket{s_k}}{_S}\\ \nonumber
&= \mathbf{I}_E
\end{align}
This implies that the quantum Fisher information of $\mathcal{E}(\rho_S)$, $\mathcal{F}(\mathcal{E}(\rho_S))$ is upper bounded by the optimal coarse grained measurement on $\rho_E$ with the same number of outcomes as the dimension on $\rho_S$ which we will denote by $F(\rho_E,G_i^*)$. Finally, we know that $F_{2|1=k'} = F(\mathcal{E}^{k'}(\ketbra{k'}),\ketbra{k})$ which implies
\begin{align}
    F_{2|1} &= \sum_{k'}q_{k'} F_{2|1=k'} \leq \sum_{k'}q_{k'} \mathcal{F}\left(\mathcal{E}^{k'}\left(\ketbra{k'}\right)\right)\\ \nonumber
    &\leq \sum_{k'}q_{k'}F(\rho_E,G_i^*)  = F(\rho_E,G_i^*).
\end{align}
This result is also interesting when the environment has a smaller dimension than the system such as might be the case in a collision model setup. In this case the Fisher information we can obtain from measuring the system is bounded by the quantum Fisher information of the environment, therefore larger probes are not necessarily more informative.

\section{Exact solution of the sequential thermometry problem}
\label{App:Thermometry}

In this appendix we give details on how to calculate the probabilities $P(k|k')$ for the metrology problem in Eqs.~\eqref{thermometry_P00}-\eqref{thermometry_Pii}. 
The system obeys the master equation~\eqref{eq::envmaster}, which forms the so-called Davies maps, which do not create coherences. Hence, the evolution after each measurement will remain diagonal and we can map this into a classical master equation problem. Define the $D$-dimensional transition matrix 
\begin{equation}
    W = \begin{pmatrix}
        -\gamma(D-1)\bar{n} & \gamma(\bar{n}+1) & \gamma (\bar{n}+1) & \ldots 
        \\
        \gamma \bar{n} & - \gamma (\bar{n}+1) & 0 & \ldots
        \\
        \gamma\bar{n} & 0 & -\gamma(\bar{n}+1) & \ldots
        \\
        \vdots & \vdots & \vdots & \ddots
    \end{pmatrix}
\end{equation}
The transition probabilities will then be 
\begin{equation}
    P(k|k') = (e^{W \tau})_{kk'},
\end{equation}
where $\tau$ is the time between measurements. 

To compute this matrix exponential we solve the corresponding master equation
\begin{equation}\label{app_thermometry_master_equation}
    \frac{dp_k}{dt} = \sum_{k'} W_{kk'} p_{k'},
\end{equation}
for all initial conditions of the form $p_{k}(0) = \delta_{k,i}$. 
We proceed in 2 steps. First, define $p_e = \sum_{i=1}^{D-1} q_i$. Then, because of the symmetry of the problem, we can actually solve a simple 2-dimensional equation for $p_0$ and $p_e$:
\begin{align}
    \frac{dp_0}{dt} &= \gamma(\bar{n}+1) p_e - \gamma(D-1)\bar{n} p_0, 
    \\
    \frac{dp_e}{dt} &= \gamma(D-1) \bar{n} p_0 - \gamma (\bar{n}+1) p_e. 
\end{align}
The solution is 
\begin{align}\label{app_thermometry_p0_solution}
    p_0(t) &= \big(q_0 + q_e f)p_0(0) + q_0(1-f) p_e(0),
    \\[0.2cm]
    p_e(t) &= q_e(1-f) p_0(0) + (q_e + q_0 f) p_e(0),
\end{align}
where $f = e^{- \gamma t(D\bar{n}+1)}$. 
From this we can already read off $P(0|0)$ in Eq.~\eqref{thermometry_P00}, as being the coefficient in Eq.~\eqref{app_thermometry_p0_solution} that multiplies $p_0(0)$. 
Similarly, because $p_e(0) = \sum_{i=1}^{D-1} p_i(0)$, we can also read off from Eq.~\eqref{app_thermometry_p0_solution} the element $P(0|i)$. 

For the remaining elements we need  to determine the probabilities of the individual states $p_i$. 
From the master equation~\eqref{app_thermometry_master_equation} we have that 
\begin{equation}
    \frac{dp_i}{dt} = - \gamma(\bar{n}+1)p_i + \gamma \bar{n} p_0(t). 
\end{equation}
Since $p_0(t)$ is known the solution will be 
\begin{align}\nonumber
    p_i(\tau) &= g p_i(0) + \gamma \bar{n} \int\limits_0^\tau dt'~g(\tau-t') \Bigg[ \big(q_0 
    \\[0.2cm]
    &+ q_e f(t'))p_0(0) + q_0(1-f(t')) (\sum_{i=1}^{D-1} p_i(0))\Bigg],
    \label{app_thermometry_pi_solution}
\end{align}
where $g\equiv g(\tau) = e^{- \gamma \tau(\bar{n}+1)}$.
The remaining matrix elements $(e^{Wt})_{kk'}$ can now all be directly read off from these results. 
For example, the element $P(i|0)$ is read from Eq.~\eqref{app_thermometry_pi_solution} by looking at all terms that multiply $p_0(0)$. 
Carrying out the time integrals we obtain the results in 
Eqs.~\eqref{thermometry_P00}-\eqref{thermometry_Pii}. 

This analysis also shows why the thermal probabilities $q_k$ are still steady-states of the Markov process, even if $\tau$ is finite. 
Namely, since $W\bm{q} = 0$, it follows that $e^{Wt}\bm{q} = \bm{q}$.

\bibliography{refs}

\end{document}